\documentclass[draft]{an}
\usepackage[round]{natbib}
\Pagespan{1}{9}      
\begin{document}
\headnote{}
\title{The criteria for a solution of the field equations to be a 
classical limit of a quantum cosmology}
\author{R. Michael Jones}
\institute{Cooperative Institute for Research in Environmental 
Sciences (CIRES)\\ University of Colorado/NOAA, Boulder, CO 
80309-0216}
\date{submitted to Astronomische Nachrichten, 3 July 2002}
\abstract{
If the gravitational field is quantized, then a solution of
Einstein's field equations is a valid cosmological model only if
it corresponds to a classical limit of a quantum cosmology.  To
determine which solutions are valid requires looking at quantum
cosmology in a particular way.  Because we infer the geometry by
measurements on matter, we can represent the amplitude for any
measurement in terms of the amplitude for the matter fields,
allowing us to integrate out the gravitational degrees of
freedom.  Combining that result with a path-integral
representation for quantum cosmology leads to an integration over
4-geometries.  Even when a semiclassical approximation for the
propagator is valid, the amplitude for any measurement includes
an integral over the gravitational degrees of freedom.  The
conditions for a solution of the field equations to be a
classical limit of a quantum cosmology are: (1) The effect of the
classical action dominates the integration, (2) the action is
stationary with respect to variation of the gravitational degrees
of freedom, and (3) only one saddlepoint contributes
significantly to each integration.}
\correspondence{R.Michael.Jones@Colorado.edu}
\maketitle
\keywords{cosmology:theory --- large-scale structure of universe}
\section{\label{Intro}Introduction}
	We normally consider all solutions of Einstein's field
equations to be valid cosological models.  However, this may not
be true if a valid cosmological model is required to be the
classical limit of a quantum cosmology.  

	Section \ref{quantum-cosmology} points out that we infer the
gravitational field from measurements on matter.  Therefore, in
comparing measurements with theory, it is sufficient to consider
the amplitudes for matter fields only, allowing us to integrate
over the gravitational degrees of freedom (an integration on a
spacelike three-dimensional hypersurface).

	Section \ref{Path-integral} points out that a path-integral
representation of the wave function involves an integration over
all 3-geometries on an initial spacelike hypersurface.  Section
\ref{4-geometries} replaces the integrations over 3-geometries on
the two spacelike hypersurfaces by the equivalent integration
over the 4-geometries connecting those two hypersurfaces.  

	Section \ref{semiclassical-approximation} considers the
semiclassical approximation for the propagator that takes the
wave function for 3-geometries and matter fields from one
spacelike hypersurface to another.  In that approximation, the
propagator depends on only one solution of the field equations. 
Solutions to the field equations fall into two categories:  
\begin{enumerate}
\item The action for the propagator dominates the behavior of the
path integral and a saddlepoint approximation is valid for each
integration in the path integral over the geometry.
\item The action for the propagator does not dominate the
behavior of the path integral or a saddlepoint approximation is
invalid for at least one of the integrations in the path integral
over the geometry.  
\end{enumerate}
Section \ref{valid} considers the former case.  Section
\ref{not-valid} considers the latter case.  Section
\ref{Interpretation} interprets these examples.\footnote{This
paper considers the meaning of quantum gravity, especially with
regard to interpreting measurements, but does not discuss
theories of quantum gravity.}

\section{\label{quantum-cosmology}Measurements in quantum
cosmology} 
	We can represent a quantum cosmology by $\left<g, \phi,
S|\psi\right>$, which is the amplitude that on a spacelike
hypersurface $S$, the 3-geometry is $g$ and the matter fields are
$\phi$.\footnote{The development here for a pure state
represented by a wave function can be generalized to a mixed
state, represented by a density matrix.}  This representation is
implicit in the path integral approach to quantum gravity
\citep{Hawking:1979}.  

	To relate this amplitude to a measurement of the geometry,
we notice that we do not measure the geometry directly.  We infer
the geometry from measurements using material objects, that is,
from measurements on the matter.  This allows us to represent any
measurement by integrating over the gravitational degrees of
freedom to give 
\begin{equation}
\left<\phi, S|\psi\right> = \int \left<g, \phi, S|\psi\right>
D(g) ,
\label{phi}
\end{equation}
the amplitude that on a spacelike hypersurface $S$, the matter
fields are $\phi$, where $D(g)$ is the measure on $g$.  

\section{\label{Path-integral}Path-integral representation}
	The wave function over 3-geometries $g_2$ and matter fields
$\phi_2$ on one 3-dimensional spacelike hypersurface $S_2$ is
related to the wave function over 3-geometries $g_1$ and matter
fields $\phi_1$ on another 3-dimensional spacelike hypersurface
$S_1$ by an extension of the path-integral 
\citep{Feynman-Hibbs:1965} formulation of quantum cosmology 
\citep{Hawking:1979} to give 
\begin{eqnarray}
&& \left<g_2, \phi_2, S_2|\psi\right> = 
\nonumber \\
&& \int \left<g_2, \phi_2, S_2|g_1, \phi_1, S_1\right> \left<g_1,
\phi_1, S_1|\psi\right> D(g_1) D(\phi_1) ,
\label{g2phi2}
\end{eqnarray}
where $\left<g_2, \phi_2, S_2|g_1, \phi_1, S_1\right>$ is the
propagator (that is, the amplitude to go from a state with
3-geometry $g_1$ and matter fields $\phi_1$ on hypersurface $S_1$
to a state with 3-geometry $g_2$ and matter fields $\phi_2$ on
hypersurface $S_2$), $\left<g_1, \phi_1, S_1|\psi\right>$ is the
wave function over 3-geometries $g_1$ and matter fields $\phi_1$
on a spacelike hypersurface $S_1$, $D(g_1)$ is the measure on
$g_1$, and $D(\phi_1)$ is the measure on $\phi_1$.  The
integration is over all 3-geometries $g_1$ and matter fields
$\phi_1$ for which the integral is defined.\footnote{Because a
matter distribution cannot be specified independent of geometry,
we take the above integration to mean that we first specify a
3-geometry, then on that 3-geometry, we specify a matter
distribution.}  

	Substituting (\ref{g2phi2}) into (\ref{phi}) gives
\begin{eqnarray}
&& \left<\phi_2, S_2|\psi\right> = 
\int \left<g_2, \phi_2, S_2|g_1, \phi_1, S_1\right> \left<g_1,
\phi_1, S_1|\psi\right> 
\nonumber \\ && 
D(g_1) D(\phi_1) D(g_2) .
\label{phi2-a}
\end{eqnarray}

\section{\label{4-geometries}Integration over 4-geometries}
	Because (\ref{phi2-a}) involves an integration over all
3-geometries $g_1$ and $g_2$ on $S_1$ and $S_2$, it is equivalent
to an integration over all 4-geometries that connect $S_1$ and
$S_2$.  Thus, (\ref{phi2-a}) can be written as
\begin{eqnarray}
&& \left<\phi_2, S_2|\psi\right> = \int \left<g_2(g^{(4)}),
\phi_2, S_2|g_1(g^{(4)}), \phi_1, S_1\right> 
\nonumber \\ && 
\left<g_1(g^{(4)}), \phi_1, S_1|\psi\right> D(g^{(4)}) D(\phi_1)
, 
\label{phi2}
\end{eqnarray}
where $D(g^{(4)})$ is the measure on the 4-geometry $g^{(4)}$.  
	Of course, until we have a full theory of quantum gravity,
we do not have formulas to give most of the functions in these
integrals.  We can, however, make some semiclassical
approximations without having a full theory.  To justify
replacing (\ref{phi2-a}) by (\ref{phi2}), we notice that the
integration in (\ref{phi2-a}) is an integration over all
4-geometries that connect $S_1$ and $S_2$, as is the integration
in (\ref{phi2}).  

\section{\label{semiclassical-approximation}semiclassical
approximation for the propagator}
	Making the semiclassical approximation\footnote{A
semiclassical approximation for the propagator is not always
valid.  Here, we consider only cases where it is valid.}  for the
propagator gives \citep{Gerlach:1969} 
\begin{eqnarray}
&& \left<g_2(g^{(4)}), \phi_2, S_2|g_1(g^{(4)}), \phi_1,
S_1\right> \approx 
\nonumber \\ && 
f_a(g_2(g^{(4)}), S_2; g_1(g^{(4)}), \phi_1, S_1) \nonumber \\ && 
e^{\frac{i}{\hbar} I_{classical} [g_2(g^{(4)}), S_2;
g_1(g^{(4)}), \phi_1, S_1] } , 
\label{prop}
\end{eqnarray}
where $I_{classical}[g_2(g^{(4)}), S_2; g_1(g^{(4)}), \phi_1,
S_1]$ is the action for the classical spacetime bounded by the
two 3-geometries that satisfies the field equations and
$f_a(g_2(g^{(4)}), S_2; g_1(g^{(4)}), \phi_1, S_1)$ is a slowly
varying function.  Explicit dependence on $\phi_2$ is not shown,
because for classical solutions to the field equations, $\phi_2$
is determined from $\phi_1$ and $g^{(4)}$.  Thus, substituting
(\ref{prop}) into (\ref{phi2})  gives
\begin{eqnarray}
&& \left<\phi_2, S_2|\psi\right> \approx 
\int f_b(g^{(4)}, \phi_1) e^{ \frac{i}{\hbar} I_{classical}
[g^{(4)}, \phi_1] } 
\nonumber \\ && 
\left<g_1(g^{(4)}), \phi_1, S_1|\psi\right> D(\phi_1) D(g^{(4)})
 .
\label{phi2class-a}
\end{eqnarray}
where $f_b(g^{(4)}, \phi_1)$ is a slowly varying function and the
integration is over all classical 4-geometries that connect $S_1$
and $S_2$.  

	The number of functions being integrated over to represent
the 4-geometry $g^{(4)}$ is probably an order of infinity greater
than that of the real numbers.  To test the validity as a
cosmology of a given 4-geometry, it is sufficient to restrict
consideration to a small subset of cases, such as a family of
known exact solutions.  This allows us to represent the
integration over 4-geometries in (\ref{phi2class-a}) more
explicitly.  Solutions to the field equations can be represented
by a number of parameters $a_i$.  These are the parameters that
specify the 4-geometry that are not constrained by the matter
distribution $\phi_1$ on the hypersurface $S_1$.  The number of
these parameters is usually finite, and in most cases, at least
countable.  I shall assume here, that they are finite, and that
there are $N$ of these parameters, although I think the
development could be extended to even the uncountable case. 
Thus, we may rewrite (\ref{phi2class-a}) more explicitly as
\begin{eqnarray}
&& \left<\phi_2, S_2|\psi\right> \approx 
\int f_c(a_i, \phi_1) e^{ \frac{i}{\hbar} I_{classical} [a_i,
\phi_1] } 
\nonumber \\ && 
\left<g_1(a_i), \phi_1, S_1|\psi\right> D(\phi_1) d^N a_i .
\label{phi2class-b}
\end{eqnarray}
where $f_c(a_i, \phi_1)$ is a slowly varying function that
depends explicitly on the parameters $a_i$ that define the
4-geometry, and now we are left with an ordinary Nth order
integral to define the integration over the 4-geometries.
\section{\label{valid}When a saddlepoint approximation is valid}
	When the behavior of $e^{\frac{i}{\hbar}I_{classical}}$
dominates over that of $\left<g_1(a_i), \phi_1, S_1|\psi\right>$
and $f_c(a_i, \phi_1)$ in the integration over each $a_i$ in
(\ref{phi2class-b}) and when a saddlepoint approximation for each
integration is valid, then we can approximate each of those
integrations by a saddlepoint approximation.  We analytically
continue each function into the complex domain, deform the path
of integration in the complex $a_i$ plane for each $a_i$ to go
through the saddlepoint, $a_{i0}$, defined by where
$I_{classical}$ is stationary for variation of each of the $a_i$,
that is, 
\begin{equation}
\left. \frac{\partial I_{classical}}{\partial a_i}
\right|_{a_i=a_{i0}} = 0 ,
\label{saddlepoint}
\end{equation}
for each $a_i$.  For each integration, the path must be deformed
(without passing over any non-analytic points) onto a steepest
descent path or a stationary phase path.  Also, to be a valid
approximation, there must not be any non-analytic points too
close to the saddlepoint.  For stationary phase paths, the
saddlepoint approximation gives e.g.
\citep{Jeffreys-Jeffreys:1978} 
\begin{eqnarray}
&& \left<\phi_2, S_2|\psi\right> \approx 
\nonumber \\ && 
\int f_c(a_{i0}, \phi_1) e^{ \frac{i}{\hbar} I_{classical}
[a_{i0}, \phi_1] } \left<g_1(a_{i0}), \phi_1, S_1|\psi\right> 
\nonumber \\ && 
e^{N i\pi/4} \prod_{i=1}^N \left|\frac{2\pi}{\partial ^2 I
/\partial a_i^2}\right|_{a_i=a_{i0}}^{1/2} D(\phi_1) .
\label{phi2class-c}
\end{eqnarray}
For steepest descent paths, the formula differs only by a phase.

	The usual form for the action $I$ is
\begin{equation}
I = \int (-|g|^{(4)})^{1/2} L d^4x , 
\label{action}
\end{equation}
where $|g|$ is the determinant of the metric tensor $g_{\mu\nu}$, 
\begin{equation}
L = \underbrace{ \frac{R-2\Lambda}{16\pi} }_{geometry} 
\underbrace{ - \frac{1}{2} \rho g_{\mu\nu} U^\mu U^\nu }_{matter} 
\underbrace{ - \frac{\rho_e}{c} A_\mu U^\mu }_{interaction} 
\underbrace{ - \frac{F_{\mu\nu} F^{\mu\nu}}{16\pi} }_{EM} 
\label{L}
\end{equation}
is the Lagrangian, $R$ is the Riemann scalar, $\Lambda$ is the
cosmological constant, $\rho$ is the mass density, $U^\mu$ is the
four-velocity, $\rho_e$ is the electric charge density, $A_\mu$
is the electromagnetic 4-vector potential, $F_{\mu\nu}$ is the
electromagnetic field tensor, and the usual designation of the
four terms is shown.\footnote{All of the examples here use
Einstein's theory of General Relativity, but the procedure
applies to nearly any gravitational theory.}

	Because the integration in (\ref{action}) must consider the
light-cone structure of the propagators, it is more appropriate
to derive a formula for the amplitude of observing a particular
event instead of deriving a general formula for all possible
measurements.  The integral for the action in (\ref{action}) must
therefore be restricted to the past light cone of the event whose
amplitude is being calculated.  There is some fuzziness to the
light cone,\footnote{There is an interesting similarity between
the light cone and the event horizon of a black hole.  In both
cases, travel across the boundary is classically possible in only
one direction (into the light cone or into the black hole), but
the prohibition is not absolute in either case, because in the
quantum situation a particle can temporarily escape by doing a
zigzag path in space-time \citep{Feynman:1962b:p.83}, resulting
in Hawking radiation in the case of a black hole.} which is taken
into account by using the correct propagators
\citep{Feynman:1962b:p.83}.

	An example of applying such a saddlepoint approximation to a
family of solutions to the field equations will be given in a
future publication.
\section{\label{not-valid}When a saddlepoint approximation is not
valid}
	We consider here several examples where the saddlepoint
approximation is either not valid or not applicable.  We take
$\Lambda$, $F_{\mu\nu}$ and $A_{\mu}$ to be zero in these
examples.  In addition, we take $R$ and $\rho$ to be zero except
where there are masses.  
\subsection{Minkowski space}
	In empty Minkowski space, the Lagrangian is everywhere zero
because the scalar curvature $R$ is zero and the matter density
is zero, and therefore, the action $I_{classical}$ is zero. 
Because there is no matter, there is no possibility for
measurements, so this case is not applicable.
\subsection{Schwarzschild metric}
	The simplest matter distribution added onto Minkowski
space-time gives us the Schwarzschild metric.  Normally, we use
the Schwarzschild metric to represent the local field around a
planet or star or black hole, but not for a whole cosmology, and
there may be good reason for that.  

	There are no gravitational degrees of freedom defining the
Schwarzschild metric, so there is no integration over
4-geometries.  However, formally, we could write
(\ref{phi2class-b}) as
\begin{eqnarray}
&& \left<\phi_2, S_2|\psi\right> \approx 
\nonumber \\ && 
\int f_c(\phi_1) e^{ \frac{i}{\hbar} I_{classical} [\phi_1] }
\left<g_1, \phi_1, S_1|\psi\right> D(\phi_1) .
\label{phi2class-d}
\end{eqnarray}
\subsection{Kerr metric}
	The next simplest model is a symmetric body like a planet
that has a rotation rate relative to an inertial frame.  We can
represent the field outside of the body by the exterior Kerr
metric.  This metric has three gravitational degrees of freedom
to characterize the direction and magnitude of the rotation rate
(which I shall refer to as $a_1$, $a_2$, and $a_3$ here). 
Because the scalar curvature and matter density are everywhere
zero outside of the body, the only contribution to the action
$I_{classical}$ is from the mass of the body, which does not
depend on the rotation rate.  Thus, (\ref{phi2class-b}) becomes
\begin{eqnarray}
&& \left<\phi_2, S_2|\psi\right> \approx 
\int e^{ \frac{i}{\hbar} I_{classical} [\phi_1] } f_d(a_1, a_2,
a_3, \phi_1) 
\nonumber \\ && 
\left<g_1(a_1, a_2, a_3), \phi_1, S_1|\psi\right> D(\phi_1) d a_1
d a_2 d a_3 ,
\label{phi2class-e}
\end{eqnarray}
where $f_d(a_1, a_2, a_3, \phi_1)$ is a slowly varying function. 
Because the exponential factor does not dominate the integration,
we cannot make a saddlepoint approximation for the integration
over $a_1$, $a_2$, and $a_3$.  We are left with an integration
over various Kerr metrics with various rotation rates.  There is
no single 4-geometry that dominates the integration.

	We normally consider the Kerr metric to represent the local
gravitational field around a spinning planet, star, or black
hole, rather than for a cosmology.  In light of the result here,
this seems appropriate.

	We want matter in the cosmological model so that we can do
measurements.  That is, because we cannot directly measure the
geometry, we must infer it from measurements on matter.  However,
the example of a single body represented here by the Kerr metric is not really interesting enough to offer the possibility for measurements of the geometry.  If we had a
planetary system, we might be able to model possible measurements
on the geometry using matter.
\subsection{\label{a-flat}Asymptotically flat metrics}
	Therefore, consider a collection of planets and a star in
some star system as the only matter in the universe.  We assume
we have some solution of the field equations for these.  In fact,
we will have many solutions, because we have some freedom in
applying boundary conditions.  

	Let us consider a subset of those solutions in which we
apply asymptotically flat boundary conditions.  Then very far
from where all of the matter is concentrated for the star system,
the solution will be approximately that of a Kerr metric, in
which the solution is characterized by the angular momentum of
the matter relative to the flat metric to which the Kerr solution
is asymptotic.  The angular momentum is characterized by 3
values, say $a_1$, $a_2$, and $a_3$.  This leads to the wave
function given by (\ref{phi2class-e}), but we cannot apply a
saddlepoint approximation because the action is independent of
$a_1$, $a_2$, and $a_3$.  
\section{\label{Interpretation}Interpretation}
	In summary, the conditions for a solution of the field
equations to be a classical limit of a quantum cosmology are: (1)
The effect of the classical action dominates the integration, (2)
the action is stationary with respect to variation of the
gravitational degrees of freedom, and (3) only one saddlepoint
contributes significantly to each integration.

	As pointed out earlier, we can always represent a
measurement of the geometry in terms of the matter; we infer the
geometry from measurements on the matter.  So, in the above
examples, what geometry would we infer from measurements on the
matter?

	Measurements on the matter in section \ref{valid} would
indicate a geometry that was confined within the limits given by
$\left|I_{saddlepoint} - I_{classical} \right| < \hbar$.  

	On the other hand, measurements on the matter in section
\ref{a-flat} would indicate an ambiguous geometry.  In fact, the
system of bodies would seem very nonclassical.  There is an
aspect of relativity here.  Although it is the background
geometry that is quantum, we can infer the geometry and matter
only relative to each other.  More specifically, we can observe
directly, only the matter, so it will appear to an observer that
the matter is behaving in a quantum manner. 

	It should be pointed out that there are no new theories or
assumptions here.  This is simply an application of standard
ideas about quantum theory to cosmology.  To falsify the results
presented here, it would be sufficient to show that our present
cosmology does not satisfy the criteria given here for a valid
cosmological model.  But unless I have made a logical error,
that would also invalidate some of our standard ideas about
quantum theory.

\end{document}